\documentclass[aps,prl,floatfix,twocolumn,amsmath,10pt,amssymb,longbibliography,superscriptaddress]{revtex4-1}

\usepackage[utf8]{inputenc}

\usepackage{graphicx}
\usepackage{stix}
\usepackage{bm}
\usepackage{hyperref}
\hypersetup{colorlinks=true, citecolor=blue, urlcolor=blue,
linkcolor=blue}
\usepackage{mleftright}
\graphicspath{{Figures/}}

\newcommand{\Dphih}{\Delta \phih}

\newcommand{\hbb}{{\bf h}_b}

\newcommand{\alphaBold}{{\bm \alpha}}

\DeclareMathSymbol{\alphaSF}{\mathalpha}{arrows2}{"0B}
\DeclareMathSymbol{\lambdaSF}{\mathalpha}{arrows2}{"15}

\newcommand{\lambdaBold}{{\bm{\lambdaSF}}}
\newcommand{\OmegaBold}{\bm \Omega}
\newcommand{\GammaBold}{\bm \Gamma}
\newcommand{\PsiBold}{\bm \Psi}
\newcommand{\OmegabBold}{\bar{\OmegaBold}}

\newcommand{\DBold}{{\bm D}}
\newcommand{\dBold}{{\bm d}}
\newcommand{\PiBold}{{\bm \Pi}}
\newcommand{\rhoBold}{{\bm \rho}}
\newcommand{\IBold}{{\bm I}}

\newcommand{\phibold}{{\bm \phi }}

\newcommand{\pb}{{\bm{p}}}
\newcommand{\qb}{{\bm{q}}}

\newcommand{\lb}{{\bf{l}}}

\newcommand{\rb}{{\bar{r}}}

\newcommand{\Gcal}{\mathcal{G}}

\newcommand{\Sigmacal}{\mathit{\Sigma}}
\newcommand{\Sigmacalh}{\Sigmacal}

\newcommand{\HF}{{\rm{HF}}}

\newcommand{\oneh}{\hat{\mathbb{1}}} 

\newcommand{\phih}{\hat{\phi}}

\newcommand{\ch}{\hat{c}}

\newcommand{\chd}{\hat{c}^\dagger}
\newcommand{\ah}{\hat{a}}
\newcommand{\ahd}{\hat{a}^\dagger}
\newcommand{\Hh}{\hat{H}}

\newcommand{\elbos}{\text{el-bos}}
\newcommand{\bos}{\text{bos}}
\newcommand{\el}{\text{el}}

\newcommand{\dif}{{\rm d}}
\newcommand{\kb}{\bm{k}}

\newcommand{\mub}{{\bar{\mu}}}
\newcommand{\nub}{{\bar{\nu}}}

\newcommand{\Eq}[1]{Eq.~\eqref{#1}}

\newcommand{\tb}{\bar{t}}

\newcommand{\alphab}{{\bar{\alpha}}}
\renewcommand{\alphab}{{\bm \alpha}}

\renewcommand{\kb}{{\bm{k}}}
\renewcommand{\lb}{{\bm{l}}}

\renewcommand{\pb}{ p}
\renewcommand{\qb}{ q}
\renewcommand{\sb}{ s}
\renewcommand{\rb}{ r}
\renewcommand{\lb}{ l}
\renewcommand{\kb}{ k}

\newcommand{\Gcalb}{{\bm \Gcal}}

\newcommand{\vBold}{{\bm v}}
\newcommand{\wBold}{{\bm w}}

\newcommand{\rhoe}{\rho}
\newcommand{\rhob}{\rhoBold_{\text{b}}}

\newcommand{\Ib}{\IBold_{\text{b}}}

\begin{document}

\title{Fast Green's function method for ultrafast electron-boson dynamics}

\author{Daniel Karlsson}
\affiliation{Department of Physics, Nanoscience Center P.O.Box 35
FI-40014 University of Jyv\"{a}skyl\"{a}, Finland}
\author{Robert van Leeuwen}
\affiliation{Department of Physics, Nanoscience Center P.O.Box 35
FI-40014 University of Jyv\"{a}skyl\"{a}, Finland}

\author{Yaroslav Pavlyukh}
\affiliation{Institut f\"{u}r Physik, Martin-Luther-Universit\"{a}t
  Halle-Wittenberg, 06120 Halle, Germany}
\affiliation{Dipartimento di Fisica, Universit\`a di Roma Tor
Vergata, Via della Ricerca Scientifica 1, 00133 Rome, Italy}

\author{Enrico Perfetto}
\affiliation{Dipartimento di Fisica, Universit\`a di Roma Tor
Vergata, Via della Ricerca Scientifica 1, 00133 Rome, Italy}
\affiliation{INFN, Sezione di Roma Tor Vergata, Via della Ricerca
Scientifica 1, 00133 Rome, Italy}

\author{Gianluca Stefanucci}
\affiliation{Dipartimento di Fisica, Universit\`a di Roma Tor
Vergata, Via della Ricerca Scientifica 1, 00133 Rome, Italy}
\affiliation{INFN, Sezione di Roma Tor Vergata, Via della Ricerca
Scientifica 1, 00133 Rome, Italy}

\begin{abstract}
The interaction of electrons with quantized phonons and photons underlies the ultrafast
dynamics of systems ranging from molecules to solids, and it gives rise to a plethora of
physical phenomena experimentally accessible using time-resolved techniques. Green’s
function methods offer an invaluable interpretation tool since scattering mechanisms of
growing complexity can be selectively incorporated in the theory. Currently, however,
real-time Green’s function simulations are either prohibitively expensive due to the {\em
  cubic} scaling with the propagation time or do neglect the feedback of electrons on the
bosons, thus violating energy conservation. We put forward a computationally efficient
Green's function scheme which overcomes both limitations. The numerical effort scales {\em
  linearly} with the propagation time while the simultaneous dressing of electrons and
bosons guarantees the fulfillment of all fundamental conservation laws. We present a
real-time study of the phonon-driven relaxation dynamics in an optically excited narrow
band-gap insulator, highlighting the nonthermal behavior of the phononic degrees of
freedom. Our formulation paves the way to first-principles simulations of electron-boson
systems with unprecedented long propagation times.
\end{abstract}

\maketitle

The time-dependent behavior of systems with strongly interacting electrons and bosons (EB)
is attracting increasing attention~\cite{Ruggenthaler2018a}.  Plasmon-polariton physics in
semiconductors~\cite{huber_how_2001,huber_femtosecond_2005,Orgiu2015}, light-enhanced
electron-phonon ($e$-$ph$) driven superconductivity~\citep{mankowsky_nonlinear_2014,
  mitrano_possible_2016,Sentef2016,babadi_theory_2017}, electron-magnon
hybridization-induced zero-bias anomalies in quantum
transport~\citep{Drewello2008,Mahfouzi2014}, manipulation of the thermoelectricity with
cavity photons~\cite{Abdullah2018} and the new field of light-driven
chemistry~\cite{Walther2006} which aims at modifying chemical reaction landscapes through
strong coupling of matter to quantized photons~\cite{Hutchison2012} is a nonexhaustive
list of possible applications. A fast and first-principles tool to deal with the quantized
nature of bosons is thus an essential requirement for future material-specific
predictions. Furthermore, such a tool may also open the way to more sophisticated
approximations of purely electronic systems, as the screened Coulomb repulsion can be
viewed as a bosonic propagator.

A full-fledged many-body method for realistic time-dependent EB systems is challenging,
however, as the quantum nature of both species has to be taken into account on the same
footing~\cite{curchod_ab_2018,kloss_multiset_2019}. Methods such as the direct solution of
the Schrödinger equation for the electron-boson wavefunction or quantum Monte Carlo
methods~\cite{Pollet2012}, scale exponentially with system size and/or
time~\cite{Cohen2015}, while other methods, such as the time-dependent matrix
renormalization group~\cite{Ma2018} are limited to model systems with a relatively small
number of basis functions. A computationally low-cost method is the extension of
time-dependent density-functional theory (DFT) to quantized
bosons~\cite{Ruggenthaler-2014,Schafer-2018,Rene-2019}, with a linear scaling in time and
a power-law scaling with system size. Nevertheless, like standard DFT, this extension
suffers from a lack of systematicity in generating approximate functionals, as well as
issues in including non-adiabatic effects.

EB interactions can instead be treated systematically through
diagrammatic~\cite{Sakkinen2015c,deMelo2016,Schuler2016} and
non-diagrammatic~\cite{Werner2013,Murakami2016,Hugel2016} expansions within the
non-equilibrium Green's function (NEGF)
formalism~\cite{Keldysh1965,Kadanoff1962,Danielewicz1984,Haug2008,Stefanucci2013,Balzer2013}. NEGF
gives access to all time-dependent one-body observables, e.g., particle density, current
density, local moments, etc., as well as to the (non)equilibrium spectral functions, and
features a power-law scaling with the size.  The main drawback of NEGF is numerical rather
than formal; the computational effort required to evolve the system by solving the
Kadanoff-Baym equations (KBE)~\cite{Keldysh1965,Kadanoff1962}\,---\,a {\em cubic} scaling
with the propagation time\,---\,limits the simulations to small systems and short times.

In purely electronic systems, the NEGF time scaling can be reduced from cubic to {\em
  quadratic} using the so-called Generalized Kadanoff-Baym Ansatz
(GKBA)~\cite{Lipavsky1986}, a controlled approximation which has recently fostered
time-dependent studies in inhomogeneous systems, from models,
~\cite{HermannsPRB2014,LPUvLS.2014,Joost2017,C60paper2018} to atoms~\cite{PUvLS.2015} and
organic molecules~\cite{PSMS.2018,PSPMS.2019,PTCNMS.2020}.  An even lower scaling has been
achieved this year, by mapping the GKBA (with mean-field propagators) integro-differential
equations onto a coupled system of ordinary differential equations (ODE). This ODE scheme
scales {\em linearly} in time~\cite{Schlunzen2020,Joost2020}, thus making NEGF a
competitor to the fastest quantum method currently available, i.e., time-dependent
DFT~\cite{Ullrich:12}. Due to a lack of an EB GKBA, however, this fast pace of progress is
confined to purely electronic systems.

This work reports on a three-fold advance of the NEGF approach to interacting EB
systems. First, we derive an EB GKBA, thereby reducing the computational effort for NEGF
EB time-propagations from cubic to quadratic. Second, we rewrite the EB GKBA
integro-differential equations as a system of ODEs, achieving \emph{time-linear} scaling
for EB systems. Third, we show that the EB GKBA scheme is conserving, i.e., the scheme
fulfills all fundamental conservation laws. These ingredients enable us to study $e$-$ph$
dynamics in an optically excited narrow band-gap insulator and to shed light on the
relaxation and nonthermal behavior of acoustic phonons.  

\paragraph{The electron-boson Hamiltonian.-}
We consider an EB system with Hamiltonian
$\Hh(t)$ given by
\begin{equation}\label{eq:totalHamiltonian}
 \Hh(t) = \Hh_\el(t) + \Hh_\bos +\Hh_\elbos(t),
\end{equation}
a sum of the electronic Hamiltonian $\Hh_\el(t)$, the bosonic one $\Hh_\bos $, and the EB
interaction $\Hh_\elbos(t)$. We do not specify $\Hh_\el(t)$, which can be any hermitian
combination of field operators $\ch_\qb$ ($\chd_\qb$) annihilating (creating) an electron
with quantum number $\qb$.  We write the free bosonic part using the displacement
$\phih_{\mu,1} \equiv \big ( \ahd_\mu + \ah_\mu \big ) / \sqrt{2}$ and the momentum
$\phih_{\mu,2} \equiv i \big ( \ahd_\mu - \ah_\mu \big )/ \sqrt{2}$, where $\ah_\mu$
($\ahd_\mu$) annihilates (creates) a boson in mode $\mu$. Introducing the composite index
$\mub = (\mu, \xi_{\mu})$ with $\xi_{\mu}=1,2$, we have
\begin{equation}
 \Hh_\bos = \sum_{\mub \nub} \Omega_{\mub \nub} \phih_\mub \phih_\nub,
\end{equation}
where \begin{math}
    \left [ \phih_{\mub},\phih_{\nub} \right ] = \alpha_{\mub \nub}
      \end{math}
and
  \begin{math}
    \alpha_{\mub \nub} = \delta_{\mu \nu}
    \begin{pmatrix}
      0  & i \\
      -i & 0
    \end{pmatrix}_{\xi_\mu \xi_\nu}.
  \end{math}
For the EB interaction we consider
\begin{equation}
 \Hh_\elbos(t) = \sum_{\mub \pb \qb} \lambda^\mub_{\pb \qb}(t)
\chd_\pb \ch_\qb \phih_\mub,
\end{equation}
with the EB coupling strength $\lambda^\mub_{\pb \qb}$. The formalism, however, is not limited
to linear coupling in the bosonic modes~\cite{marini_functional_2018}.

\paragraph{The electron-boson KBE.-}
In the NEGF formalism the fundamental unknowns are the electronic lesser/greater
$G^{\lessgtr}$ single-particle Green's function (GF) and the bosonic counterparts, $D^{\lessgtr}$. They satisfy the KBE, a system of nonlinear integro-differential equations which for the electronic part read (in matrix form):
\begin{align}\label{eq:ElectronicKBE}
\begin{split}
\Big[ i \overrightarrow{\partial}_{t} - h (t) \Big ] G
^\lessgtr(t,t')
\! = \!
\left [\Sigmacalh^\lessgtr \cdot G^A + \Sigmacalh^R \cdot
G^\lessgtr \right]\!(t,t'),
\\
G ^\lessgtr(t,t')
\Big [ -i \overleftarrow{\partial}_{t'} -h(t') \Big ]
\! = \!
\left [G^\lessgtr \cdot \Sigmacalh^A + G^R \cdot
\Sigmacalh^\lessgtr  \right ]\! (t,t'),
\end{split}
\end{align}
where
$
\left [A \cdot B\right ] (t,t') \equiv \int \dif \tb\,
A(t,\tb) B(\tb,t'),
$
is a real-time convolution and
\mbox{
$
 X^{R/A}(t,t') = \pm \theta[\pm(t-t')] \left [ X^>(t,t') - X^<(t,t')
\right ]\!
$
}
is the retarded/advanced function.  The quantity $\Sigmacalh$ is the correlation part of
the self-energy, whereas the time-local mean-field part is incorporated in the
single-particle Hamiltonian $h(t) = h_\HF(t)+ h_\bos(t) $, where $h_\HF(t)$ is the
Hartree-Fock Hamiltonian and $h_{\bos,\pb \qb}(t)=\sum_{\mub} \lambda^\mub_{\pb \qb}(t)
\phi_\mub(t)$ is the bosonic potential.  The expectation value $\phi_\mub(t) = \langle
\phih_{H,\mub}(t)\rangle$ ($H$ denotes the Heisenberg picture) fulfills in matrix form
\begin{equation}\label{eq:expValuePhi}
 \left[i\alphab \frac{\dif}{\dif t} - \OmegabBold\right]\phibold(t) =
  \sum_{\pb \qb} \lambdaBold_{\pb \qb}(t) \rhoe_{\qb \pb}(t).
\end{equation}
In Eq.~(\ref{eq:expValuePhi})  $\OmegabBold \equiv \OmegaBold +
\OmegaBold^T$ and $\rhoe(t) \equiv \rhoe^{<}(t)= -i G^<(t,t)$ is the electronic
single-particle density matrix.

\begin{figure}[tbp]
  \centering
\includegraphics[width=0.9\columnwidth]{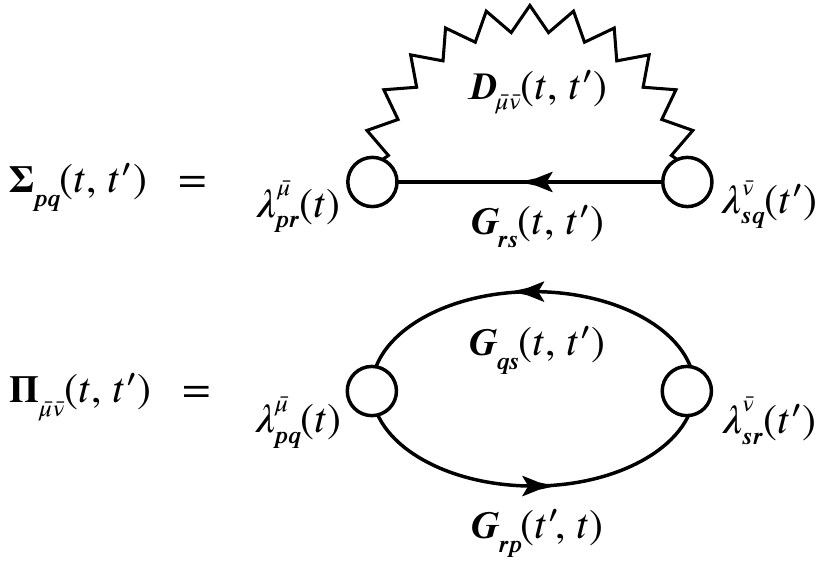}
 \caption{The $G\DBold$ approximation for the electronic (upper) and random phase
   approximation for the bosonic (lower) self-energies.}
 \label{fig:diag2}
\end{figure}

The bosonic GFs are defined using the fluctuation
operators $\Dphih_{H,\mub}(t) = \phih_{H,\mub}(t) -
\phi_{\mub}(t)$:
\begin{equation}\label{eq:DlessGreat}
 D^<_{\mub \nub}(t,t')  = -i \big\langle \Dphih_{H,\nub}(t')
\Dphih_{H,\mub} (t)  \big\rangle,
\end{equation}
and $D^>_{\mub \nub}(t,t') = D^<_{\nub \mub}(t',t)$.
The expectation value of $\Dphih_{H,\mub}(t)$ is
identically zero by construction, a property which simplifies the
bosonic KBE~\cite{Sakkinen2015c,Karlsson2018b}:
\begin{align}\label{eq:bosonicKBE}
\begin{split}
  \left [i \overrightarrow{\partial}_t - \alphaBold \OmegabBold \right ]\!
\DBold^\lessgtr(t,t') =
  \alphaBold\!
 \left [
   \PiBold^\lessgtr \cdot \DBold^A
   +
   \PiBold^R \cdot \DBold^\lessgtr
 \right]\!(t,t'),
 \\
 \DBold^\lessgtr(t,t')\! \left [-i
   \overleftarrow{\partial}_{t'} -  \OmegabBold\alphaBold \right ]
 \! = \!
 \left [
   \DBold^\lessgtr \cdot \PiBold^A
   +
   \DBold^R \cdot \PiBold^\lessgtr
 \right]\!(t,t')\alphaBold,
\end{split}
 \end{align}
where $\PiBold$ is the bosonic self-energy. In the $\phi$-field notation, the bosonic KBE
are first-order in time.  The numerical solution of the coupled
Eqs.~(\ref{eq:ElectronicKBE}) and (\ref{eq:bosonicKBE}) is demanding (cubic scaling with
the number of time steps) and so far achieved only in small model
systems~\cite{Sakkinen2015a,Sakkinen2015c,Schuler2016,Karlsson2018b}. In this work, we
consider the $G \DBold$ approximation shown diagrammatically in Fig.~\ref{fig:diag2}, as
well as the $G \dBold$ ($\PiBold = 0$) and mean-field ($\Sigma = \PiBold =0$, also known
as semi-classical Ehrenfest) approximation.

\paragraph{The electron-boson GKBA.-}
The KBE can be used to generate an equation of motion (EOM) for the electronic density
matrix $\rhoe(t)$ and its bosonic counterpart $\rhob(t) \equiv \rhob^<(t)= i
\DBold^<(t,t)$. As $\rhoe(t)$ and $\rhob(t)$ are single-time functions, their calculation
scales quadratically with the number of time steps. Subtracting the two equations in
\Eq{eq:ElectronicKBE} and \Eq{eq:bosonicKBE} and then letting $t'\to t$ yields
\begin{align}
    \label{electronicRho}
    \begin{split}
    &\partial_t \rhoe(t) + i \left [ h (t),  \rhoe(t) \right ]  = -
    \left(I(t) + I^\dagger(t) \right),
    \\
    &\partial_t \rhob(t)
  +i \left [\alphaBold \OmegabBold \rhob(t) - \rhob(t)\OmegabBold
\alphaBold \right ]  =
 \Ib(t) + \Ib^T(t),
 \end{split}
\end{align}
with the electronic and bosonic collision integrals defined as
\begin{align}\label{eq:electronicCollision}
 \begin{split}
 I (t)  &=
 \int_{0}^t \! \! \dif \tb \left [\Sigmacalh^>(t,\tb)
G^<(\tb,t) - \Sigmacalh^<(t,\tb) G^>(\tb,t) \right],
\\
\Ib(t)
  &=
 \alphaBold \!\int _0^t \! \dif \tb
 \left [\PiBold ^>(t,\tb) \DBold^<(\tb,t)
  - \PiBold ^<(t,\tb)\DBold^>(\tb,t) \right].
\end{split}
\end{align}
Evaluation of the collision integrals requires the time-off-diagonal lesser and greater
GFs; hence \Eq{electronicRho} is not a closed system of equations for the density
matrices. A partial rescue is provided by the electronic GKBA~\cite{Lipavsky1986}, i.e.,
$G^\lessgtr(t,t') = \mp \left[G^R(t,t') \rhoe^\lessgtr(t') - \rhoe^\lessgtr(t)G^A(t,t')
  \right]$, where $\rhoe^>(t) \equiv \oneh - \rhoe(t)$.  Taking $ G^R(t,t') = -i
\theta(t-t') \mathcal{T}\mleft\{\exp\mleft(-i \int_{t'}^{t} h (\tb)
d\tb\mright)\mright\}, \label{Gret} $ and $G^A(t,t') = [G^{R}(t',t)]^\dagger$ at the
mean-field level, the lesser/greater electronic GF's become functionals of $\rhoe^{<}(t)$.
However, to close \Eq{eq:electronicCollision}, a GKBA-like form of the lesser/greater
bosonic GF is needed.

The form of the electronic GKBA is motivated by the fulfillment of the
mean-field KBE, but is augmented with a correlated density
matrix. Using the same argument we have derived the
\emph{bosonic} GKBA~\cite{SuppMat}
\begin{equation}\label{bosonicGKBA}
 \DBold^\lessgtr(t,t') =  \DBold^R(t,t') \alphaBold
\rhob^\lessgtr(t') - \rhob^\lessgtr(t) \alphaBold \DBold^A(t,t'),
\end{equation}
where $\rhob^>(t) = \alphaBold + \rhob(t)$.  Taking $\DBold^{R/A}(t,t') = \mp i \alphaBold
\theta[\pm(t-t')] e^{-i \OmegabBold \alphaBold (t-t')}$ at the mean-field level (which
coincides with the noninteracting case~\cite{Sakkinen2015c}) the lesser/greater bosonic
GF's become functionals of $\rhob^{<}(t)$. The bosonic GKBA in \Eq{bosonicGKBA} applies
even if $\OmegabBold$ depends explicitly on time (e.\,g., phonon
driving~\cite{forst_nonlinear_2011})~\cite{SuppMat}. The EB GKBA allows for closing the system in \Eq{electronicRho}
as both collision integrals $I$ and $\Ib$ become functionals of $\rhoe$ and $\rhob$.
Together with the equation for $\phi$, \Eq{eq:expValuePhi}, the dynamics of any EB system
can be simulated.

\paragraph{\it Conservation laws.-}
The EB GKBA scheme is conserving, i.e., all fundamental conservation laws are fulfilled
provided that the underlying diagrammatic approximation to $\Sigma=\Sigma[G,\DBold]$ and
$\PiBold=\PiBold[G,\DBold]$ stem from the functional derivatives of the Baym functional
$\Phi[G,\DBold]$~\cite{Baym1962} (for the EB case, see, for example,
\cite{Sakkinen2016,Tokatly2018}).  Although Baym's original derivation pertains to
self-consistent solutions of the KBE the whole proof goes through if the r.h.s.  of
Eqs.~(\ref{eq:ElectronicKBE}) and (\ref{eq:bosonicKBE}) are evaluated at GF's $G'$ and
$\DBold'$ (and hence at $\Phi$-derivable self-energies $\Sigma'=\Sigma[G',\DBold']$ and
$\PiBold'=\PiBold[G',\DBold']$) different from the GF's $G$ and $\DBold$ appearing in the
l.h.s..  In Supp.~Mat~\cite{SuppMat} we show that conservation laws are recovered up to
terms proportional to the change of $\Phi[G',\DBold']$, as $G'$ and $\DBold'$ are changed
according to the transformation having the conserved quantity as generator. Since $\Phi$
is invariant under these special transformations the aforementioned terms vanish.  In the
context of particle conservation this fact was pointed out in Ref.~\cite{Mera2012} for
$G'$ the one-shot GF of an electronic system. The argument is, however, more general and
holds for all conservation laws, including energy conservation, as well as EB systems,
thereby enlarging enormously the class of conserving approximations.

As the $G\DBold$ self-energy is $\Phi$-derivable and the GKBA approximation for $G$ and
$\DBold$ is one out of the infinitely many choices for $G'$ and $\DBold'$, our scheme is
fully conserving and, in particular, it correctly balances the energy transfer from
electrons to bosons and viceversa. The $G\dBold$ approximation instead is not
$\Phi$-derivable, bosons do not feel any feedback from the electrons, and energy
conservation is jeopardized.

\paragraph{Linear-time scaling of the electron-boson GKBA.-}
The EB GKBA computational cost scales quadratically with the number of time steps, as the
domain of integration for $I(t)$ and $\Ib(t)$ grows linearly in time.  Remarkably, the
time-scaling can be further reduced from quadratic to linear {\em without affecting the
  scaling with the system size}. Let us write the collision integrals of
\Eq{eq:electronicCollision} in the $G\DBold$ approximation as
\begin{align}\label{eq:CollIntGCal}
    \begin{split}
 I_{\pb \lb}(t)
 &=
 i \sum_{\mub \rb } \lambda^\mub_{\pb \rb}(t)
 \Gcal^{\mub}_{\rb \lb}(t),
\\
 \Ib (t)
 &=
 -i\sum_{\rb \lb}
  \left[\alphaBold   \lambdaBold_{\rb \lb} (t)\right]
 \otimes \Gcalb_{\lb \rb}(t),
 \end{split}
\end{align}
where we introduced the tensor product $(\vBold \otimes \wBold)_{\mub \nub} = v_\mub
w_\nub$ and the one-time vector   $ \Gcalb_{\rb \lb} =
\Gcalb^{>}_{\rb \lb} - \Gcalb^{<}_{\rb \lb}$ with
\begin{equation}\label{eq:bigG}
 \Gcalb^{\lessgtr}_{\rb \lb}(t)
 = \sum_{\sb \qb}
 \int _{0}^t \dif \tb \ \DBold^\lessgtr(t,\tb) G^\lessgtr_{\rb
\sb}(t,\tb) \lambdaBold_{\sb \qb}(\tb)  G^\gtrless_{\qb \lb} (\tb,t).
\end{equation}
Differentiating \Eq{eq:bigG} with respect to time yields
\begin{align}
\begin{split}\label{eq:difEqBigG}
  i\frac{\dif}{\dif t} \Gcalb_{\rb \lb}(t)
  &=
 \PsiBold_{\rb \lb}(t)
+
\alphaBold \OmegabBold  \Gcalb_{\rb \lb}(t)
\\
&\quad
+
\sum_\kb
  \left[
    h_{\rb \kb}(t) \Gcalb_{\kb \lb}(t) - \Gcalb_{\rb \kb}(t)h_{\kb
\lb}(t)
  \right],
\end{split}
\end{align}
with $\Gcalb_{\rb \lb}(t=0) = 0$,
$\PsiBold_{\rb \lb}(t) =  \PsiBold_{\rb \lb}^>(t)  -
\PsiBold_{\rb \lb}^<(t)$, and
\begin{equation}
 \PsiBold_{\rb \lb}^\lessgtr(t) =
 \rhob^\lessgtr(t) \sum_{\sb \qb}
   \rhoe^\lessgtr_{\rb \sb}(t) \lambdaBold_{\sb \qb}(t)
\rhoe^\gtrless_{\qb \lb} (t).
\end{equation}
In obtaining \Eq{eq:difEqBigG} we used the Leibnitz rule
of differentiation,
\begin{math}
 \frac{\dif}{\dif t} \Big ( \int_{0}^t \dif \tb \ f(t,\tb)\Big)
 =
 f(t,t) + \int_{0}^t \dif \tb \ \frac{\partial}{\partial t}
f(t,\tb),
\end{math}
and the fact that the GKBA GF's satisfy the mean-field KBE, i.e.,
$i\frac{\partial}{\partial t} G^\lessgtr(t,t') = h(t) G^\lessgtr(t,t')$ and $
i\frac{\partial}{\partial t} \DBold^\lessgtr(t,t') = \alphaBold \OmegabBold
\DBold^\lessgtr(t,t')$. The equations for $\rhoe$, $\rhob$, \Eq{electronicRho}, and
$\Gcalb$, \Eq{eq:difEqBigG}, form a closed system of first-order ODEs which is {\em
  equivalent} to the original EB GKBA integro-differential equations.  Since no
integration over time is needed, the EB ODE scheme scales linearly in time.

\paragraph{Numerical algorithms.-}
We have numerically checked that the integro-differential and ODE formulations of the EB
GKBA yield the same results, up to numerical accuracy.  We implemented the former scheme
in the \textsc{Cheers}~\cite{Perfetto2018b} code. The algorithm for the bosonic case
follows the electronic algorithm closely, with the difference that the time-propagation is
non-unitary as $\alphaBold$ and $\OmegabBold$ do not commute. However, by defining the
hermitian matrices $\hbb = \frac{1}{2}(\alphaBold \OmegabBold + \OmegabBold \alphaBold)$
and $\GammaBold = \frac{i}{2}(\alphaBold \OmegabBold - \OmegabBold \alphaBold)$, inserting
them into the bosonic EOM and absorbing $\GammaBold$ into the collision integral, the
bosonic equation gets the same structure as the electronic one and can be solved using the
same algorithm. The linear-time propagation is done using the fourth-order Runge-Kutta
solver.  In Supp.~Mat.~\cite{SuppMat}, we provide numerical evidence of the performance
and accuracy of the method in the paradigmatic Holstein model, a hallmark of strongly
interacting EB systems. The EB GKBA is benchmarked against exact results as well as full
numerical solution of the EB Kadanoff-Baym equations, finding a satisfactory agreement
even in the strong-coupling regime.  The scaling with the system size is determined by two
parameters: The dimension of the electronic basis, $N_\text{e}$, and the number of bosonic
modes, $N_\text{b}$. We emphasize that the method does not scale with the number of
electrons or bosons. In particular, the scaling is $\mathcal{O}(N_\text{e}^3 \times
N_\text{b})$ and $\mathcal{O}(N_\text{e}^2 \times N_\text{b}^3)$ for computing electronic
$I$ and bosonic $\Ib$ collision integrals, respectively.

\paragraph{Two band model.-}
\begin{figure}[t!]
  \centering \includegraphics[width=0.8\columnwidth]{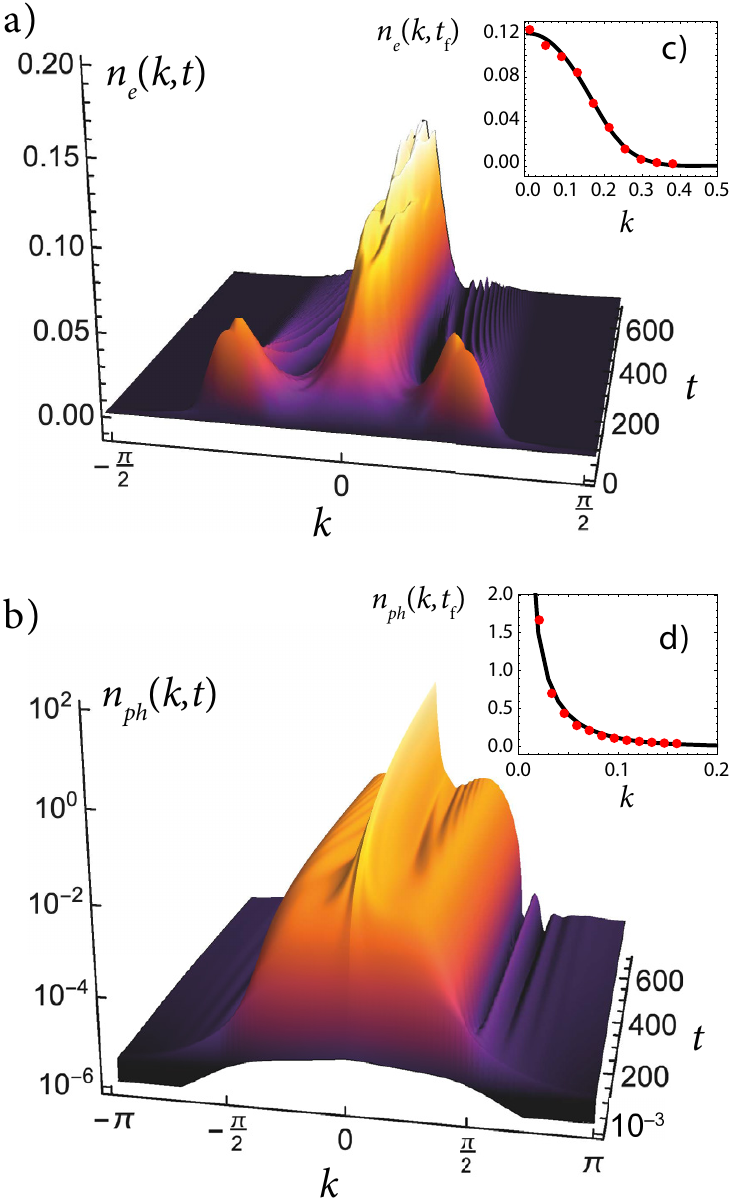}
 \caption{Relaxation of conduction electrons (a) and acoustic phonons (b) in a two-band
   model excited with a laser pulse of optical frequency $\omega_P=1.4$, Rabi frequency
   $\Omega_P=0.06$, and pulse duration $T_P=20$ (energies and times in units of
   $\varepsilon_g$ and $\varepsilon^{-1}_g$, respectively).  Electrons are coupled to a
   branch of acoustic phonons with $\omega_D=0.8$ and coupling strength
   $\lambda_{cc}=0.05$. Insets (c) and (d) depict $e$ and $ph$ populations at the end of
   the propagation. They can be well-fitted with the Fermi-Dirac and the Bose-Einstein
   distributions at inverse temperature $\beta=80$.}
 \label{fig:thermalize}
\end{figure}

To demonstrate the capabilities of our method we turn to periodic systems, specifically a
model of a narrow band-gap insulator consisting of one valence $v$ and one conduction $c$
band~\cite{perfetto_floquet_2020}.  Due to system's translational invariance the momentum
representation is appropriate:
\begin{align}
  \hat{H}_\text{el}&=\sum_{\alpha, k}\varepsilon_{\alpha k}\hat{c}_{\alpha
    k}^\dagger\hat{c}_{\alpha k} +\frac{1}{N_k}\!\sum_{q,k,k'} U_{q}^{cv}\hat{c}_{c
    k+q}^\dagger\hat{c}_{v k'-q}^\dagger\hat{c}_{v k'}\hat{c}_{c k}.
\end{align}
Here $U_{q}^{cv}$ is the Fourier transform of the interband soft Coulomb interaction
$U_{ij}^{cv}=U/\sqrt{|i-j|^2+1}$ and $N_k$ is the number of $k$-points. $e$-$e$
interaction is treated at the mean-field level. The electron dispersion
$\varepsilon_{\alpha k}$ is described by two parameters: the bandwidth $W$ and the band
gap $\varepsilon_g$. Henceforth we express all energies in units of $\varepsilon_g$ and
choose $U=W/2=1$~\cite{perfetto_pump-driven_2019,perfetto_floquet_2020}. The electronic
system is coupled to a single ($\mu=1$) phononic branch
\begin{align}
  \hat{H}_\text{bos}&=\sum_{q}\omega_{\mu q}\hat{a}^\dagger_{\mu q}\hat{a}_{\mu q},\quad\omega_{\mu q}=\frac{\omega_D}{\pi}|q|;\\
  \hat{H}_\text{el-bos}&=\sum_{k,q}\lambda^{\mu}_{cc}(k,q)\hat{c}_{c k}^\dagger\hat{c}_{c k}(\hat{a}^\dagger_{\mu q}+\hat{a}_{\mu q}).
\end{align}
The coupling is momentum-independent,
$\lambda^{\mu}_{cc}(k,q)=\lambda_{cc}$. We consider \emph{acoustic}
phonons with linear dispersion characterized by the Debye frequency $\omega_D$ at the edge
of the Brillouin zone $q=\pm\pi$ (in units of inverse lattice spacing). Initially the
system is in the ground state, hence the conduction band and the phonons are not
populated, $n_{e}(k)=n_{ph}(k)=0$. We solve the EB GKBA equations using a mesh of
$N_k=1500$ points. In the $k$-space formulation the scaling with the system size reduces
to $\mathcal{O}(N_k^2)$, see Supp. Mat.~\cite{SuppMat}.

In Fig.~\ref{fig:thermalize} we present the $e$-$ph$ dynamics triggered by a laser pulse
of frequency $\omega_P=1.4$. Because $\omega_P>\varepsilon_g$, the $c$-band is populated at
nonzero momentum $\pm k_0$ (see the two domes at $t\simeq 0$ in panel a). With electrons
in the $c$-band, the $e$-$ph$ scattering becomes relevant, leading to the creation of phonons, the
subsequent redistribution of $n_e(k)$ and $n_{ph}(k)$ occupations and, eventually, to the
thermalization of the electrons and low-momentum phonons as well as to the generation and
re-absorption of nonthermal phonons~\cite{yang_observation_2015} around the $\pm k_0$
hot-spots.  For a typical value of the gap $\varepsilon_g=1.1$\,eV the frequency
$\omega_P$ corresponds to the 800\,nm wave-length of Ti-sapphire laser. The kinetic energy
of the conduction electrons immediately after the pulse is then $0.22$\,eV yielding for
the inverse temperature $\beta=2.9\,\text{eV}^{-1}$\footnote{By assuming parabolic
dispersion of the conduction electrons at small-$k$ it follows
$E_K=-\zeta_{3/2}/[2^{3/2}\zeta_{1/2}\beta]=0.63/\beta$} or approximately
$T_e=4000$\,K. At the end of propagation electrons and low-momentum phonons are
thermalized, see insets (c,d) in Fig.~\ref{fig:thermalize}, with $\beta\approx80$,
corresponding in our example to $T_e\approx T_{ph}=160$\,K. Signatures of the initially
hot phonon-distribution do instead persist for much longer times, as can be seen from
the side bands at approximately $\pm k_0$. The intermediate stages of the dynamics are
more complex, they are characterized by at least two time-scales (associated with
$\omega_D$ and $\lambda_{cc}^2/\omega_D$) describing the rapid creation of the
nonequilibrium phonons and their slow thermalization.

This wealth of phenomena cannot be observed in simpler approaches, such as those based on
the two-temperature model~\cite{shin_extended_2015}, semiclassical Boltzmann transport
equation~\cite{sadasivam_theory_2017}, or even in NEGF theories with frozen
phonons~\cite{kemper_direct_2015,sangalli_ultra-fast_2015,murakami_ultrafast_2020}.  The
coupled $e$-$ph$ dynamics can be studied using the nonequilibrium dynamical mean field
theory (DMFT)~\cite{freericks_nonequilibrium_2006}. However, in this scheme nonlocal
correlations are difficult to
incorporate~\cite{biermann_first-principles_2003}. Furthermore, DMFT applications have
been so far limited to the Hubbard-Holstein
model~\cite{murakami_interaction_2015,murakami_nonequilibrium_2017,murakami_multiple_2016}
with optical phonons, which simplies the momentum treatment. Here we demonstrate that it
is possible to consider realistic $e$- and $ph$-dispersions and do the propagation
linearly in time. Applications to light enchanced
superconductivity~\cite{forst_nonlinear_2011, mankowsky_nonlinear_2014,
  babadi_theory_2017}, formation and melting of the excitonic
orders~\cite{hellmann_time-domain_2012}, ultrafast band gap
control~\cite{mor_ultrafast_2017} and many other emerging light-induced
phenomena~\cite{basov_towards_2017} are envisaged.

{\it Conclusions.-}
We have derived an EB GKBA approximation for bosonic propagators and put forward a NEGF
scheme to simulate the correlated dynamics of EB systems.  The formal advantages of the
methods are (i) approximations can be systematically improved by a proper selection of
Feynman diagrams and (ii) all fundamental conservation laws are fullfilled provided that
the self-energy diagrams are $\Phi$-derivable. The energy conservation makes the EB GKBA
suitable for studying a plethora of situations where electrons and, for example, phonons
can exchange energy; our example being carrier relaxation in a pumped insulator system
with acoustic phonons.  The computational effort of solving the EB GKBA equations in the
$G\DBold$ approximation scales linearly in time; they can also be implemented in more
advanced diagrammatic approximations using the same strategies outlined in
Ref.~\cite{Joost2020}.  The inclusion of $e$-$e$ interactions in the linear-scaling
scheme, as discussed in Refs.~\cite{Schlunzen2020,Joost2020}, is straightforward.  We
therefore believe that our proposed method provides an efficient and accurate alternative
to the existing computational tools for models as well as first-principles simulations of
interacting electrons and bosons out of equilibrium.

\begin{acknowledgments}
D.K. likes to thank the Academy of Finland for funding under Project No. 308697.
R.v.L. likes to thank the Academy of Finland for support under grant no. 317139.  G.S, E.P
and Y.P. acknowledge the financial support from MIUR PRIN (Grant No. 20173B72NB), from
INFN through the TIME2QUEST project, and from Tor Vergata University through the Beyond
Borders Project ULEXIEX.
\end{acknowledgments}

\end{document}